\documentclass[10pt]{iopart}

\usepackage{iopams}
\usepackage{graphicx}
\newcommand{\ket}[1]{|{#1}\rangle}
\newcommand{\bra}[1]{\langle {#1}|}

\begin{document}

\title{Generalized entanglement as a framework for complex quantum
systems: Purity vs delocalization measures  \ }

\author{Lorenza Viola and Winton G. Brown}

\address{Department of Physics and Astronomy, Dartmouth
College, 6127 Wilder Laboratory, Hanover, NH 03755, USA}

\ead{Corresponding author:lorenza.viola@dartmouth.edu}

\begin{abstract}
We establish contact between the delocalization properties of pure
quantum states, as quantified by their number of principal components,
and the average generalized entanglement properties, as quantified by
purity measures relative to different observable sets.  We find that
correlations between products of state vector components with respect
to Hamming distance play an important role in the structure of
subsystem-based purity measures.  In particular, we derive general
conditions under which the amount of global multipartite entanglement
relates to the inverse participation ratio averaged over a maximal set
of mutually unbiased product bases.  Furthermore, we provide a method
for computing the expected amount of generalized entanglement with
respect to an arbitrary observable set for random pure states.
Specific examples and an explicit application to a disordered quantum
spin chain are discussed.
\end{abstract}

\submitto{\JPA}

\section{Introduction}

Developing a deeper qualitative and quantitative understanding of
``complex'' quantum systems is a broad challenge whose implications
range from condensed matter physics to fundamental quantum theory and
quantum information science (QIS).  In a loose sense, complexity may
be intuitively associated with the lack of a ``simple'' description of
physical properties in situations where such a description should {\em
in principle} follow from a small set of known, basic rules
\cite{Michael}.  In quantum systems, complex quantum features so
defined may appear at both the kinematical and dynamical level via
three main pathways: large state-space size; interaction between
constituent subsystems; absence of dynamical regularities and
non-integrability.  Taken together, these factors may be ultimately
held responsible for non-scalable (typically, exponentially
inefficient) parameterizations of system properties; the emergence of
many-body phenomena and quantum irreversibility (via interactions both
within the system of interest and between the system and its
environment); and the possibility of dynamical instability and quantum
chaos.

Physically, the occurrence of genuinely quantum correlations --
entanglement -- lies at the heart of all the above phenomena.  From a
QIS perspective, entanglement is also intimately tied to the inherent
complexity that distinguishes quantum from purely classical
information processing -- being, in particular, the defining resource
for quantum communication \cite{QComm} as well as a necessary
prerequisite for quantum computational speed-up \cite{QComp}.  In
recent years, this has naturally motivated a host of investigations
aimed at characterizing the nature and role of entanglement in complex
quantum systems.  While a complete understanding is far from being
reached, important progress is being made toward elucidating the
behavior of entanglement across a quantum phase transition
\cite{QPT,GEQPT} and across a transition from integrability to quantum
chaos \cite{Qchaos,Mejia,Baker}, as well as in better accounting for
entanglement in computational schemes for interacting quantum systems,
like renormalization group methods \cite{RN}.

In this context, the notion of {\em generalized entanglement} (GE)
\cite{GE} has recently emerged as a unifying framework for describing
entanglement in {\em arbitrary} physical and QIS settings -- capable,
in particular, to recover conventional (subsystem-based) entanglement
in well-defined conditions and to directly incorporate physical
constraints such as quantum indistinguishability.  Beside providing
new Lie-algebraic measures for diagnosing broken-symmetry quantum
phase transitions in a variety of models \cite{GEQPT}, GE has
contributed so far to the understanding of standard multipartite spin
correlations in disordered lattice systems \cite{Monta}, provided a
natural testbed for investigating entanglement generation in chaotic
quantum maps \cite{QKT}, as well as shed light on conditions for
quantum-computational speed-up in a wide class of Lie-algebraic models
\cite{Somma06}.

Here, we continue to explore applications of the GE framework to
complex quantum systems by focusing, in particular, on highlighting
the relationship between GE measures and {\em state delocalization
properties}, as quantified by standard indicators like the {\em number
of principal components} with respect to an appropriate basis.  In
line with the growing body of work at the interface between QIS,
condensed-matter, and quantum statistical physics \cite{RMP}, our main
motivation is to qualitatively and quantitatively characterize points
of contact between notions originally developed in different contexts
-- in the hope that this may lead to useful cross-implications.
Following a review of the essential GE background in Section 2, the
relationship between global multipartite entanglement and
delocalization is addressed in Section 3, by explicitly uncovering the
role of Hamming distance in the structure of subsystem-based GE.  In
Section 4, a general method for estimating the expected GE of {\em
random pure states} is presented.  In Section 5, a concrete
application to a disordered Heisenberg spin-$1/2$ chain is discussed.
Final remarks conclude in Section 6.

\section{Generalized entanglement and purity measures}

The basic idea of the GE approach is that the entanglement properties
of quantum states are determined by the expectation values of a {\em
distinguished subspace of observables} rather than a preferred
decomposition of the system into subsystems \cite{GE}.  This allows
the notion of GE to be meaningful in physical settings which the
conventional notion is too narrow to embrace: in particular, GE is
directly applicable to systems subject to limitations in the available
control interactions and measurements, and described by arbitrary
operator languages (spin, fermion, etc) as for instance many-body
quantum systems.  In addition, because the GE notion rests only on
convexity properties of spaces of quantum states and observables, GE
is mathematically suited for entanglement formulations in abstract
operational theories.  We refer the reader to
\cite{GE,CMT,GEBS,Cones,VB} for a more expanded discussion.

The starting point for defining GE is to realize that a pure state of
a composite quantum system is entangled (in the usual sense) iff at
least one of the reduced subsystem states is mixed.  Let the system of
interest be described by a pure state, $|\psi\rangle \in {\cal H}$ ,
with ${\cal H}$, dim$({\cal H})=N$, and $\rho = \ket{\psi}\bra{\psi}$
being the associated Hilbert space and density operator, respectively.
Let in addition the distinguished observable set consists of the
Hermitian operators in a linear subspace $h \subseteq {\cal B}({\cal
H})$ of the full operator space on ${\cal H}$, with $h$ closed under
Hermitian conjugation.  The key step is to replace the notion of
reduced state as obtained via a partial trace in the usual
tensor-product sense by a notion of reduced state as resulting from
the restriction to $h$ of the positive linear functional $\omega$
corresponding to $\rho$ via the trace map \cite{GE,GEBS,Cones}.  Such
a reduction may be specified in terms of the (unique) projection map
${\cal P}_h$ with respect to the trace inner product, $\rho \mapsto
{\cal P}_h(\rho)$.  Accordingly, $|\psi\rangle$ is defined to be {\em
generalized unentangled relative to $h$} iff its reduced state ${\cal
P}_h(\rho)$ is pure -- that is, extremal in the space of reduced
states \cite{remark}.

While no unique measure suffice to quantify the amount of $h$-GE
present in $|\psi\rangle$, the simplest possibility is to evaluate the
square length of ${\cal P}_h(\rho)$.  Let $\big\{b_i\big\}$ be a basis
of Hermitian traceless operators for $h$, orthogonal in the trace
inner product, $\mbox{tr}(b_i b_j) = N \delta_{ij}$.  The {\em purity
of $|\psi\rangle$ relative to $h$} ($h$-purity) is given by
\begin{equation}
P_h (|\psi\rangle )= \kappa_h \sum_i \mbox{tr}(\rho b_i)^2 =\kappa_h
\sum_i |\langle \psi|b_i |\psi\rangle|^2,
\label{Ph}
\end{equation}
where the normalization constant $\kappa_h$ depends in general on $h$
and $N$, and ensures that the maximum value of $P_h$ is 1.  Thus, a
state $\ket{\psi}$ with maximal purity, $P_h(\ket{\psi})=1$, is
unentangled with respect to $h$, hence it has extremal length
\cite{Cones}.  In the physically relevant case where $h$ forms a
(irreducibly represented) Lie algebra, maximal $h$-purity is both a
necessary and sufficient condition for a pure state to be
$h$-unentangled, $P_h$ is invariant under unitary transformations
generated by arbitrary elements of $h$, and GE$_h \equiv 1-P_h$ is an
entanglement monotone \cite{GE}.  $P_h$ may be extended to a measure
for mixed-state GE via a standard convex-roof construction \cite{GE}.

The following specializations and applications of the above GE
definition may serve to clarify the relationship between GE and
standard entanglement, and will be especially relevant for the present
discussion:

\vspace*{1mm}

(1) {\em Absolute purity}.  By definition, $\rho$ is pure iff it is a
one-dimensional projector, hence iff $\mbox{tr}(\rho^2)=1$. By
identifying $h$ with the (real) Lie algebra of {\em all} traceless
observables on ${\cal H}$, $h=\mathfrak{su}(N)$, Eq.~(\ref{Ph}) gives
$\kappa_{all} =1/(N-1)$ and
\begin{equation}
P_{all} (|\psi \rangle ) = P_{\mathfrak{su}(N)} (|\psi\rangle ) =
\frac{N}{N-1} \Big(\mbox{tr}(\rho^2) - \frac{1}{N} \Big),
\label{fullP}
\end{equation}
consistently normalized so that purity is 1 for pure states and 0 for
totally mixed ones.

\vspace*{1mm}

(2) {\em Bipartite systems and linear entropy}. For a system
consisting of two subsystems $A$ and $B$,
$\mathcal{H}=\mathcal{H}_A\otimes \mathcal{H}_B$, dim$({\cal
  H}_A)=d_A$, dim$({\cal H}_B)=d_B$, the information accessible
through measurements on $A$ or $B$ alone is contained in the reduced
density operators $\rho_A = \mbox{tr}_B(\rho)$, $\rho_B =
\mbox{tr}_A(\rho)$.  As mentioned, a pure state $\ket{\psi}\in
\mathcal{H}$ is unentangled, $\ket{\psi} =
\ket{\phi_A}\otimes\ket{\phi_B}$, iff both $\rho_A$ and $\rho_B$, are
pure.  A bipartite entanglement measure known as the linear entropy
$E$ (of either subsystem) may be constructed as $E_A (|\psi\rangle) =
1- \mbox{tr}(\rho_A^2)$.  In the GE approach, standard bipartite
entanglement is recovered by choosing the set of {\em all} uni-local
observables acting on $A$ (or $B$) alone, e.g. $h=h_{A}=
\mathfrak{su}(d_A)\oplus \mathbb{I}_B$ (equivalently,
$e^{ih}=\mbox{SU}(d_A) \otimes {\mathbb I}_B$).  Eq.~(\ref{Ph}) then
yields
\begin{equation}
1-P_{h_A} (|\psi\rangle) = 1 - \frac{d_A}{d_A-1} \Big( \mbox{tr}(\rho_A^2) -
\frac{1}{d_A} \Big) = \frac{d_A}{d_A-1} E_A (|\psi\rangle),
\end{equation}
that is, GE$_{h_A}$ is directly proportional to the linear subsystem
entropy.

\vspace*{1mm}

(3) {\em Multipartite systems and average subsystem purity}.  The
above example generalizes to a multipartite system consisting of $n$
subsystems of dimension $d$.  That is, conventional (subsystem-based)
entanglement is recovered by selecting the algebra of all uni-local
observables acting on individual subsystems as distinguished
observables, $h=h_{loc}= \oplus_i h_i = \mathfrak{su}_1(d)\oplus
\ldots \oplus \mathfrak{su}_n(d)$.  Eq.~(\ref{Ph}) then gives
\begin{equation}
P_{loc} (|\psi\rangle) = \frac{d}{d-1} \Big[ \frac{1}{n} \sum_{i=1}^n
\Big( \mbox{tr}(\rho_i^2) - \frac{1}{d} \Big) \Big]
= \frac{1}{n} \sum_{i=1}^n P_{h_i} (|\psi\rangle),
\end{equation}
that is, the purity with respect to arbitrary local observables is
equal to the average (normalized) subsystem purity, as intuitively
expected \cite{GEBS}.  $P_{loc}(|\psi\rangle)$ attains its maximum 1
only for completely separable states, $\ket{\psi} =\otimes_{i=1}^n
\ket{\phi_i}$, and is equal to 0 iff each reduced density matrix is
totally mixed (hence no information is available through local
operations).  The entanglement measure GE$_{loc}=1-P_{loc}$ is thus
proportional to the average linear entropy over all bi-partitions of
the system into blocks of $1$ and $(n-1)$ subsystems.  For qubit
systems ($d=2$), such a measure has been shown in \cite{GEQPT} and
\cite{Brennen} to coincide with global multipartite entanglement $Q$
as introduced by Meyer-Wallach \cite{MW}, $Q(|\psi\rangle) =
1-P_{loc}(|\psi\rangle)$.

\vspace*{1mm}

(4) {\em Expected $h$-purity of a set of states}.  For a fixed
observable set, the expected $h$-purity of a pure state taken with
respect to a certain probability distribution $\xi$ quantifies GE
properties of a {\em typical} state in the ensemble,
\begin{equation}
\overline{P}_h = {\mathbb E}^{(\xi)}\{  P_h (|\psi\rangle) \}\,.
\label{expP}
\end{equation}
An important instance arises for uniformly sampled random pure
states, in which case $\xi$ coincides with the unitarily invariant
Haar measure on $\mbox{SU}(N)$ \cite{Zycz}.

\section{Delocalization and local purity}

Given an orthonormal basis $\{\ket{k}\}$ in ${\cal H}$, a
well-established measure of state delocalization in quantum
statistical physics and quantum chaos is the {\em number of principal
components} (${\tt NPC}$),
\begin{equation}
{\tt NPC}(\ket{\psi}) = \Big( \sum_k |\langle k \ket{\psi}|^4
\Big)^{-1} = \Big( \sum_k |a_k |^4 \Big)^{-1} , \hspace{5mm}
\sum_k |a_k|^2=1\,,
\label{npc}
\end{equation}
quantifying the number of basis states on which $|\psi\rangle$ has a
significant amplitude $a_k \in {\mathbb C}$.  {\tt NPC} so defined
ranges from a minimum value of 1, meaning that $|\psi\rangle$
coincides with a single basis element, to a maximum of $N$,
corresponding to a maximally delocalized state with equal
probabilities $|a_k|^2=1/N$.  If ${\tt NPC}>1$, measurements in the
corresponding basis will result in a probability distribution over
possible outcomes.  For instance, a crossover from localization to
delocalization with respect to a large number of basis states occurs
in the eigenvectors of the Anderson model during the
insulator-to-metal transition, as well as in the eigenvectors of
quantum spin lattices across a transition to quantum chaos
\cite{Qchaos}. {\tt NPC} is equivalently referred to as {\em
participation ratio} (or participation number \cite{Mejia}).
Accordingly, ${\tt NPC}^{-1}$ will be often denoted here as {\em
inverse participation ratio} ({\tt IPR}) \cite{Haake,note}.

From a conceptual standpoint, it is interesting to observe that ${\tt
NPC}$ (and {\tt IPR}) may be directly related to an appropriate
$h$-purity.  Specifically, let $h=h_{diag}$ denote the subspace of all
(traceless) observables which are diagonal in the chosen orthonormal
basis $\{\ket{k}\}$. Then,
\begin{equation}
P_{h_{diag}} (|\psi\rangle) =\frac{N}{N-1}\frac{1}{{\tt NPC}
(|\psi\rangle)} - \frac{1}{N-1}.
\label{Pdiag}
\end{equation}
Such an observable space may be considered the commutant of a
non-degenerate Hamiltonian and forms a trivial, abelian Lie algebra.
As such $h_{diag}$ does not identify a decomposition into quantum
subsystems, and GE$_{diag}$ need not have any relationship to
entanglement in the standard sense.  Clearly, $P_{h_{diag}}
(|\psi\rangle)=1$ iff ${\tt NPC} (|\psi\rangle)=1$. Thus, in a sense,
$P_{h_{diag}} (|\psi\rangle)$ may be also thought as quantifying how
non-classical $|\psi\rangle$ is relative to the given basis.

Our next objective is to investigate to what extent a relation similar
to Eq.~(\ref{Pdiag}) may exist between standard entanglement (as
quantified by $P_{loc}$) and ${\tt NPC}$, as evaluated in each of a
maximal set of mutually unbiased product bases \cite{Jay}.  A product
basis is one where each basis state is unentangled.  Two bases are
mutually unbiased if localization in one basis implies maximal
delocalization in the other. In general, we shall find that $P_{loc}$
is not solely a function of ${\tt NPC}$ in these bases, but it also
depends on additional structure of the input state.

Focus on a system consisting of $n$ qubits (spin-$1/2$) first,
$N=2^n$. The bases $\big\{\ket{k_\alpha}\big\}$, consisting of the
joint eigenstates of qubit observables $\big\{\sigma_\alpha^{(i)}\big
\}$, $\alpha =x,y,z$, provide a natural maximal set of mutually
unbiased product bases. Let $\big\{a_k^\alpha\big\}$ denote the
components of $\ket{\psi}$ in the basis $\big\{\ket{k_\alpha}\big\}$.
The local purity $P_{loc}$ may then be expressed as $P_{loc} = P_x +
P_y + P_z$, where $P_{\alpha}(|\psi\rangle) = \frac{1}{n} \sum_i
\bra{\psi} \sigma_\alpha^{(i)} \ket{\psi}^2$ \cite{GE,GEBS}. Recall
that the {\em Hamming distance} between two binary bit strings of
equal length measures the number of substitutions required to change
one into the other.  We have:

\vspace*{1mm}

{\sc Lemma 3.1}. For every pure state $|\psi\rangle$ of $n$ qubits,
the following identity holds:
\begin{equation}
P_{\alpha}(\ket{\psi}) = 1- \frac{4}{n}\sum_{k<j}f_{kj}
|a_k^\alpha|^2|a_j^\alpha|^2 ,
\label{flips}
\end{equation}
where $f_{kj}$ is the Hamming distance between basis states
$\ket{k_\alpha}$ and $\ket{j_\alpha}$ that is, the number of
instances where the eigenvalues of $\sigma_\alpha^{(i)}$ differ on
$\ket{k_\alpha}$ and $\ket{j_\alpha}$.

\vspace*{1mm}

{\sc Proof.} Note that we may express
$$\bra{\psi}\sigma_\alpha^{(i)}\ket{\psi} = \sum_k |a_k^\alpha|^2 -
\sum_{k'} |a_{k'}^\alpha|^2,$$ \noindent where the (un)primed sum
is over all $a_k^\alpha$ such that the $k$-th basis state has a
(0)1 for the $i$-th qubit.  Squaring both sides yields
\begin{eqnarray}
\bra{\psi}\sigma_\alpha^{(i)}\ket{\psi}^2 &=&\sum_k
|a_k^\alpha|^4 +\sum_{k'}|a_{k'}^\alpha|^4 \label{longeq} \\
&+& 2 \Big( \sum_{k<j}
|a_k^\alpha|^2|a_j^\alpha|^2 + \sum_{k'<j'} |a_{k'}^\alpha|^2
|a_{j'}^\alpha|^2 - \sum_{kk'} |a_k^\alpha|^2|a_{k'}^\alpha|^2\Big).
\nonumber
\end{eqnarray}
From the normalization of $|\psi\rangle$ one obtains
\begin{equation}
\sum_k|a_k^\alpha|^4 \equiv {\tt IPR}_\alpha =
1-2\sum_{k<j}|a_k^\alpha|^2|a_j^\alpha|^2.
\label{norm}
\end{equation}
Substituting (\ref{norm}) into (\ref{longeq}) yields
$$\bra{\psi}\sigma_\alpha^i\ket{\psi}^2 = 1 -
4\sum_{kk'}|a_k^\alpha|^2|a_{k'}^\alpha|^2.$$ \noindent Note that this
sum is over all pairs such that $\ket{k_\alpha}$ and $\ket{k'_\alpha}$
differ on the $i$th qubit.  Hence, the number of occurrences of a term
$|a_k^\alpha|^2|a_j^\alpha|^2$ in the sum yielding $P_\alpha$, which
is over all qubits, is equal to the Hamming distance between
$\ket{k_\alpha}$ and $\ket{j_\alpha}$, whereby the result.\hfill$\Box$

\vspace*{1mm}

Note the structural similarity between the expressions for $P_\alpha$
and ${\tt IPR}_\alpha$, Eqs.~(\ref{flips}) and (\ref{norm}) -- the
main difference being that in $P_\alpha$ the products
$|a_k^\alpha|^2|a_j^\alpha|^2$ are weighted by Hamming distance
whereas in ${\tt IPR}_\alpha$ they are not.  We have the following:

\vspace*{1mm}

{\sc Theorem 3.1}. Assume that for each basis $\alpha=x,y,z$, the
values of the terms $|a_k^\alpha|^2|a_j^\alpha|^2$ are independent
on average upon Hamming distance ({\em uncorrelation assumption}).
Then for every pure state $|\psi\rangle$ of $n$ qubits,
\begin{equation}
P_{loc} (|\psi\rangle) = \Big(\frac{N}{N-1} \sum_{\alpha=x,z,y} {\tt
IPR}_\alpha (|\psi\rangle) \Big) - \frac{3}{N-1},
\label{no dep}
\end{equation}
where $N = 2^n$ is the dimension of the Hilbert space.

\vspace*{1mm}

{\sc Proof}. Let $A_f^\alpha =
\overline{|a_k^\alpha|^2|a_j^\alpha|^2}$ denote the average over all
pairs $k,j$, constrained to a specific Hamming-distance value $f_{kj}
= f$, and let $A^\alpha = \overline{|a_k^\alpha|^2|a_j^\alpha|^2}$
denote the unconstrained average over all $k$ and $j$.  Then the
weighted average over all pairs $k,j$ may be separated into the sum of
averages over pairs corresponding to a given $f$,
$$ \sum_{k,j} f_{kj} |a_k^\alpha|^2|a_j^\alpha|^2 = \sum_{f} n_ff
A_f^\alpha = \sum_f n_f f \Big(\frac{A_f^\alpha}{A^\alpha}\Big)
A^\alpha\equiv \Big( \sum_f n_f f w_f^\alpha \Big) A^\alpha ,$$
\noindent
where $n_f$ is the number of pairs $k,j$ with fixed Hamming distance
$f$. Under the uncorrelation assumption, each of the ratios
$w_f^\alpha =1$, irrespective of $f$.  Thus, by invoking the
expression of $P_\alpha$ in Lemma 3.1, and by making the average over
pairs defining $A^\alpha$ explicit,
\begin{eqnarray*}
P_\alpha (|\psi\rangle) &= &1-\frac{4}{n}\Big( \sum_{f} n_f f w_f^\alpha
\Big)\Big(\frac{2}{N(N-1)}\sum_{k<j}
|a_k^\alpha|^2|a_j^\alpha|^2\Big)\\
& = & 1-
\frac{4}{nN(N-1)}\Big(\sum_{f}n_f  f \Big) \Big( 1-{\tt
IPR}_\alpha (|\psi\rangle)\Big).
\end{eqnarray*}
\noindent
To evaluate $\sum_f n_f f$, first note that for each state, $\ket{k}$,
there are ${n \choose f}$ states labelled by $j$ that are Hamming
distance $f$ from $\ket{k}$. Thus, $n_f = \frac{N}{2} {n\choose
f}$. Using ${n\choose f} = {n\choose n-f}$, it follows that
$\sum_{f=0}^n f {n\choose f} =\frac{n}{2}\sum_{f=0}^n {n\choose f} =
\frac{n}{2}N$. Hence, $\sum_f n_f f = \frac{n N^2}{4}$. By summing
over $\alpha$, the result follows. \hfill$\Box$

\vspace*{1mm}

Thus, $P_{loc}$ depends in general on both the ${\tt NPC}$ in a set of
three mutually unbiased product bases and on the {\em average
correlation of the products $|a_k|^2|a_j|^2$ with respect to Hamming
distance} in each basis.

\subsection{Conditions for single-basis delocalization}

For states obeying certain symmetries, the number of bases involved in
the relationship between delocalization and $P_{loc}$ may be reduced.

A first physically relevant example is provided by states invariant
under a non-standard (anti-unitary) time reversal symmetry $T$ such
that $T^2 =\mathbb{I}$ \cite{Haake}. All states invariant under $T$
may be expressed using only {\em real} components in an appropriate
basis. For such states the expectation values involving the imaginary
part of the operator space are zero.  For instance, for states
$\ket{\psi}$ of qubit systems which are real in the standard
$\big\{\ket{k_z}\big\}$ basis, it follows that
$\bra{\psi}\sigma_y^{(i)}\ket{\psi} = 0$ for all $i$.  Hence, $P_y =
0$, and ${\tt NPC}_y$ does not enter the expression for $P_{loc}$.

Notably, a further simplification occurs for the energy eigenstates of
a large class of two-body spin Hamiltonians, which includes the
Heisenberg, XXZ, XY, and Ising models -- specifically, any Hamiltonian
which may be written in the form
$$H= \sum_i \varepsilon_i \sigma_z^{(i)} + \sum_{i,j}
J_z^{(i,j)}\sigma_z^{(i)}\sigma_z^{(j)} +
J_x^{(i,j)}\sigma_x^{(i)} \sigma_x^{(j)} +
J_y^{(i,j)}\sigma_y^{(i)} \sigma_y^{(j)}\,,$$
\noindent
for arbitrary coupling parameters $J^{(i,j)}_\alpha \in {\mathbb R}$
and on-site energy splittings $\varepsilon_i \in {\mathbb R}$. Any
such Hamiltonian commutes with the collective Pauli operator
$\bigotimes_{i=1}^n \sigma_z^{(i)}$, which describes a global $Z_2$
symmetry.  If $H$ is {\em non-degenerate}, then each eigenvector is
invariant under this symmetry.  It then follows from standard
properties of Pauli operators (namely, that if $[\sigma_a,\sigma_b]
\ne 0$ and $\sigma_a \ket{\psi} = \lambda_a \ket{\psi}$, then
$\bra{\psi}\sigma_b\ket{\psi} = 0$) that
\begin{equation}
\bra{\psi}\sigma_x^{(i)}\ket{\psi} =
\bra{\psi}\sigma_y^{(i)}\ket{\psi} = 0,~\; \forall i\,.
\label{xy}
\end{equation}
Under such conditions, $P_{loc}= P_z$, hence global entanglement
properties, depend only on ${\tt NPC}_z$ and Hamming-correlations in
the $\big\{\ket{k_z}\big\}$ basis \cite{caveat}.

Note that if $\varepsilon_i = 0 \:, \forall i\,$, then
$\bigotimes_{i=1}^n \sigma_x^{(i)}$ and $\bigotimes_{i=1}^n
\sigma_y^{(i)}$ are also symmetries of $H$, resulting in
$\bra{\psi}\sigma_z^{(i)}\ket{\psi} = 0,\; \forall i\,$, thereby
yielding $P_{loc} \equiv 0$ for each eigenvector of $H$.

\vspace*{1mm}

An important class of states obeying Eq.~(\ref{xy}) are the
eigenstates of total $z$-angular momentum, $S_z = \sum_i
\sigma_z^{(i)}$. For the $S_z = 0$ subspace, for instance, the
assumption of no correlation between $A_f^z$ and $f_{kj}$ implies
\begin{equation}
P_{loc}^{(S_z=0)}(|\psi\rangle) = \frac{N_0}{N_0-1} \frac{1}{{\tt NPC}_z
(|\psi\rangle)} - \frac{1}{N_0-1},
\label{no dep Sz0}
\end{equation}
where $N_0 = {n!}/{[(n/2)!]^2}$ is the dimension of the subspace.  In
the $n$-dimensional $S_z={n}-{2}$ subspace describing the
single-excitation sector, $f_{kj}=2$ for all pairs of basis states,
hence $P_{loc}$ depends directly on ${\tt NPC}_z$,
\begin{equation}
P_{loc}^{(S_z=n-2)} (|\psi\rangle) = \frac{4}{n} \frac{1}{{\tt NPC}_z
(|\psi\rangle)} + \frac{n-4}{n}\,,
\label{Szn2}
\end{equation}
in agreement with the relationship between average linear entropy and
delocalization of one-particle states found in \cite{Wang}.

\vspace*{1mm}

{\sc Remark 3.1.} It may be interesting to observe that, for an {\em
arbitrary} state $\ket{\psi}$, it is always possible to identify a
product basis where
\begin{equation}
P_{loc}(\ket{\psi}) = 1-\frac{4}{n}\sum_{k<j}f_{kj}
|a_k|^2|a_j|^2. \label{flips2}
\end{equation}

To show this, note that the expectation value of an arbitrary
traceless normalized $i$th-qubit observable may be written as
$n_x^{(i)} \bra{\psi}\sigma_x^{(i)}\ket{\psi} + n_y^{(i)}
\bra{\psi}\sigma_y^{(i)}\ket{\psi} + n_z^{(i)}\bra{\psi}
\sigma_z^{(i)}\ket{\psi}$, for some unit vector
$\vec{n}^{(i)}=(n_x^{(i)},n_y^{(i)},n_z^{(i)})$. Since the 3-tuple of
real numbers $(\langle\sigma_x^{(i)}\rangle$,
$\langle\sigma_y^{(i)}\rangle$, $\langle\sigma_z^{(i)}\rangle$) can be
considered a vector in $\mathbb{R}^3$, there clearly exists a
direction $\vec{n}^{(i)}$ that is parallel to the vector of
expectations.
One may associate a traceless normalized single-qubit operator with
the parallel direction for each qubit, $\tilde{\sigma}_z^{(i)}=
\vec{n}^{(i)}\cdot\vec{\sigma}^{(i)}$. The mutual eigenstates of the
$\{\tilde{\sigma}_z^{(i)}\}$ then form a product basis in which
Eq.~(\ref{flips2}) holds since the single-qubit operators
perpendicular to the $\{\tilde{\sigma}_z^{(i)}\}$ have vanishing
expectation values. In this basis, the reduced density matrix of each
subsystem is diagonal.  As such, the expression for a state in this
basis may be considered a standard canonical form which generalizes
(non-uniquely) the Schmidt decomposition for bipartite systems
\cite{Sudbery,Todd}.

\subsection{Generalization to qudit systems}

Eq.~(\ref{flips2}) and, under appropriate conditions, Theorem 3.1,
may be generalized to a system consisting of $n$ $d$-dimensional
subsystems (qudits).

To this end, begin by observing that because each reduced density
matrix is Hermitian, it is always possible (similar to the $d=2$ case)
to find a product basis where each qudit reduced density matrix is
diagonal.  Let each state in such a basis be specified in terms of
quantum numbers $\ket{v_1,\ldots ,v_n}$, where $v_i$ labels a state of
the $i$-th qudit and may take any of $d$ possible values. Let
$\ket{\psi} = \sum_k a_k\ket{k}$, where $k$ is a collective index
ranging over all possible strings of values $(v_1,\ldots,v_n)$.  The
reduced density matrix for the $i$-th qudit may then be expressed as
$\rho_i = \sum_{v_i} \big( \sum_{k|k_i =v_i}
|a_k|^2\big)\ket{v_i}\bra{v_i}$, and $\rho_i^2 = \sum_{v_i} \big(
\sum_{k,k'| k_i= k'_i=v_i} |a_k|^2|a_{k'}|^2 \big)
\ket{v_i}\bra{v_i}$, whereby $\mbox{tr}(\rho_i^2)$ is the sum over all
terms $|a_k|^2|a_{k'}|^2$ such that $k_i=k'_i$. Hence a term
$|a_k|^2|a_{k'}|^2$ occurs in the sum over different qudits, $\sum_i
\mbox{tr}(\rho_i^2)$, as many times as $v_i = v'_i$. Let $f_{kk'}$ be
the number of instances in which $v_i\neq v'_i$ over all $i$. This may
be considered a {\em generalized Hamming distance}, which reduces to
the usual one for $d=2$. Now $\sum_i \mbox{tr}(\rho_i^2) =
\sum_{kk'}(n-f_{kk'})|a_k|^2|a_{k'}|^2$.  Using the identity
$\sum_{kk'} |a_k|^2|a_{k'}|^2 = 1$, this may be rewritten as $\sum_i
\mbox{tr}(\rho_i^2) = n - \sum_{kk'}f_{kk'}|a_k|^2|a_{k'}|^2$.  Thus,
\begin{equation}
P_{loc} (|\psi\rangle) = 1 -
\frac{2d}{n(d-1)}\sum_{k<k'}f_{kk'}|a_k|^2|a_{k'}|^2\,.
\label{flipsgen}
\end{equation}

When the value of $d$ is such that a maximal set of $(d+1)$ mutually
unbiased bases spanning the state space of each qudit exists, the
construction in \cite{Bandy} implies the existence of an Hermitian
operator basis for the unilocal observables on each qudit which is
partitioned into $(d+1)$ maximally commuting subsets.  Accordingly,
the local purity may be written as $P_{loc} = \sum_{\alpha=1}^{d+1}
P_\alpha$, where $P_\alpha$ is the purity with respect to a choice of
one maximally commuting set of basis operators for each of the
qudits. The operators contributing to each $P_\alpha$ uniquely define
(up to irrelevant relabeling transformations) a product basis, $\big\{
\ket{k_\alpha} \big\}$, where $k_\alpha$ is a collective index for the
local quantum numbers, $v_\alpha^i$, which label the mutually unbiased
bases of each qudit.  Since for any $\alpha$ and $\beta$
$|\bra{v_\alpha^1 ... v_\alpha^n} v_\beta^1 ... v_\beta^n \rangle|^2 =
|\bra{v_\alpha^1} v_\beta^1 \rangle ...  \bra{v_\alpha^n} v_\beta^n
\rangle|^2 = 1/d^n$ for all values of $\big\{v_\alpha^i\big\}$ and
$\big\{v_\beta^i\big\}$, it follows that the product bases $\big
\{\ket{k_{\alpha}}\big\}$ are mutually unbiased.

Recall that Eq.~(\ref{flipsgen}) results from summing the squared
diagonal matrix elements of each reduced density operator in a
particular product basis. The stipulation that each reduced density
matrix is diagonal in this basis ensures that such a sum yields
$P_{loc}$ (after subtracting the trace contribution and proper
normalization). If such a condition is relaxed, then
Eq.~(\ref{flipsgen}) states a relationship between the purity with
respect to an operator basis spanning the diagonal observables of each
qudit and the components of state vectors along the corresponding
product basis.  Hence Eq.~(\ref{flipsgen}) will in general hold
between each $P_\alpha$ and the state vector components in the
corresponding basis $\big\{\ket{k_\alpha}\big\}$. Considerations
similar to the ones presented for the qubit case are then applicable.
Thus, $P_{loc}$ may be related in general to ${\tt NPC}$, and
correlations with respect to (generalized) Hamming distance in each
mutually unbiased product basis.

\vspace*{1mm}

{\sc Remark 3.2.} Interestingly, the expression
$\sum_{k<k'}f_{kk'}|a_k|^2|a_{k'}|^2$ may be interpreted as the
expectation value of the (generalized) Hamming distance between
measurements on two copies of the state $\ket{\psi}$. Thus, the local
GE, GE$_{loc}= 1-P_{loc}$, may always be written as
\begin{equation}
\mbox{GE}_{loc} (|\psi\rangle) = \frac{2d}{n(d-1)}
\bra{\psi}\otimes\bra{\psi}\,F\,\ket{\psi}\otimes{\ket\psi},
\end{equation}
where $F$ is a Hermitian operator which may be interpreted as the
Hamming distance between measurements on two copies of the same state
in the canonical, state-dependent basis in which the subsystem reduced
density matrices are diagonal.

For $d$ such that each qudit may be spanned by each of a maximal set
of mutually unbiased bases, $\mbox{GE}_{loc}$ may additionally be
expressed in terms of the expectation values of Hamming distance
between measurements of two copies of the same state in each of
$(d+1)$ mutually unbiased product bases,
\begin{equation}
\mbox{GE}_{loc}(|\psi\rangle) = \frac{2d}{n(d-1)}
\bra{\psi}\otimes\bra{\psi}\,\Big( \sum_{\alpha=1}^{d+1} F_\alpha
\Big) \,\ket{\psi}\otimes{\ket\psi} - d\,.
\end{equation}

\section{Average generalized entanglement of random pure states}

The requirement of Hamming uncorrelation which is responsible for a
simple relationship between global multipartite entanglement and
delocalization is naturally satisfied on average by certain classes of
random states.

One such family may be defined, for instance, by taking an arbitrary
set of normalized probabilities, assigning them at random to basis
states in $\big\{\ket{k_z}\big\}$, and giving each component a random
phase.  The resulting ${\tt NPC}_z$ is determined exactly by the set
of probabilities, and is the same for all states in the ensemble. The
distribution of components in the $\big\{\ket{k_x}\big\}$ and
$\big\{\ket{k_y}\big\}$ bases, and hence the expected value of ${\tt
NPC}_x$ and ${\tt NPC}_y$ is determined by the set of
probabilities. For particular states of the ensemble ${\tt NPC}_x$ and
${\tt NPC}_y$ will fluctuate around this value. The random assignment
of probabilities ensures that in the $\big\{\ket{k_z}\big\}$ basis no
correlation between component products and Hamming distance exists on
average. Furthermore, the random phases ensure Hamming uncorrelation
in the $\big\{\ket{k_x}\big\}$ and $\big\{\ket{k_y}\big\}$ bases also.
Thus, ensemble averages over many random assignments will yield the
relationship in Eq.~(\ref{no dep}).

In practice, random states generated by uniformly sampling according
to the invariant Haar measure play an important role, naturally
emerging, in particular, within statistical descriptions of complex
many-body systems such as Random Matrix Theory (RMT) \cite{RMT}.
Results on the expected linear entropy of a subsystem date back to
early work by Lubkin \cite{Lubkin}, have been further extended in
\cite{Page}, and more recently revisited in the context of obtaining
estimates of the expected value and variance of the Meyer-Wallach
global entanglement \cite{Baker}, and generalizations to other
bipartite divisions \cite{Scott}.  Results on the full probability
distribution have also been established under additional restrictions
on the set of states and/or entanglement measure
\cite{Patrick,Plenio,Facchi}.  Here, we begin investigating {\it
typical GE properties} with respect to an {\em arbitrary} observable
set, and show that a simple method allows to calculate the expected
$h$-purity, $\overline{P_h}$, defined in Eq.~(\ref{expP}). We have the
following:

\vspace*{1mm}

{\sc Theorem 4.1.} Let $h$ be any (Hermitian closed) subspace of
traceless observables on ${\cal H}$.  The expected $h$-purity of a
pure state sampled uniformly according to the Haar measure is given by
\begin{equation}
\overline{P}_h = {\mathbb E}^{(Haar)}\{{P_h} (|\psi\rangle)\} =
\kappa_h \frac{\mbox{dim}(h)}{N+1}.
\label{ranPh}
\end{equation}

\vspace*{1mm}

{\sc Proof.}  We first show that the ensemble expectation
${\mathbb E}\{{\bra{\psi}b_i\ket{\psi}^2}\}$ is the same for any
normalized traceless operator spanning $h$.  Let $b_i =
\sum\lambda_i\ket{n}\bra{n}$ be a spectral decomposition of $b_i$.
Now, ${\mathbb E}\{{\bra{\psi}b_i\ket{\psi}^2} \}= \sum\lambda_n^2
{\mathbb E}\{ |\langle{n}\ket{\psi}|^4 \} +2\sum\lambda_n\lambda_m
{\mathbb E}\{ |\langle{n}\ket{\psi}|^2|\langle{m}\ket{\psi}|^2
\}.$ Since by assumption the distribution of $\ket{\psi}$ is
invariant under arbitrary unitary transformations, the expectation
${\mathbb E}\{ |\langle{n}\ket{\psi}|^4 \}$ is the same for all
$n$, and ${\mathbb E}\{
|\langle{n}\ket{\psi}|^2|\langle{m}\ket{\psi}|^2 \}$ is the same
for all pairs $m\neq n$.  From the trace and normalization
conditions, $\sum\lambda_n = 0$, and $\sum\lambda_n^2 = N$, it
follows that $ \sum\lambda_m\lambda_n = -{N}/{2}$.  Thus,
${\mathbb E} \{ \bra{\psi}b_i\ket{\psi}^2 \} = N {\mathbb E} \{
|\langle{0}\ket{\psi}|^4 \} - N {\mathbb E} \{
|\langle{0}\ket{\psi}|^2|\langle{1}\ket{\psi}|^2 \}$, irrespective
of $i$.

The value of ${\mathbb E}\{ \bra{\psi}b_i\ket{\psi}^2 \}$ may be
determined by using the property that the purity relative to the full
space of observables equals 1.  Since, by Eq.~(\ref{fullP}),
$\kappa_{all}=1/(N-1)$, and $(N^2-1)$ linearly independent traceless
operators exist, the required expectation is
\begin{equation}
{\mathbb E}\{ \bra{\psi}b_i\ket{\psi}^2 \} = \frac{1}{\kappa_{all} (N^2-1)}
= \frac{1}{(N+1)}.
\label{bav}
\end{equation}
\noindent
The expected $h$-purity is $\overline{P_h} = \kappa_h\sum_i {\mathbb
E}\{ \bra{\psi}b_i\ket{\psi}^2 \}$, which yields the desired
result.\hfill$\Box$

\vspace*{1mm}

{\sc Example 1.} For a system of $n$ qubits, the local purity of a
typical pure state averaged over the Haar measure on SU($2^n$) is
found to be
$$ \overline{P}_{loc}= \kappa_{loc} \frac{3n}{N+1} = \frac{3}{N+1},$$
\noindent in agreement with the result for $\overline{\mbox
GE}_{loc}=\overline{Q}= (N-2)/(N+1)$ derived in \cite{Baker}.

\vspace*{1mm}

{\sc Example 2.} As a further application, consider a spin-$J$ system,
living in a Hilbert space of dimension $N=2J+1$, carrying an
irreducible representation of SU(2). If SU(2) observables are
distinguished, the corresponding $\mathfrak{su}(2)$-purity is
$$ P_{\mathfrak{su}(2)}(|\psi\rangle) = \frac{J+1}{3J}
\sum_{\ell=x,y,z} \langle \psi |b_\ell| \psi \rangle^2 ,\hspace{5mm}
b_\ell=\sqrt{\frac{3}{J(J+1)}}\,J_\ell, $$ where $J_\ell$ denote
angular momentum operators, and $\kappa_{\mathfrak{su}(2)}=(J+1)/3J$
is chosen so that $P_{\mathfrak{su}(2)} (|\psi\rangle) = 1$ for
angular momentum generalized coherent states \cite{Arecchi}.  The
above GE measure may be directly relevant to describe GE generation in
a quantum kicked top initially prepared in a spin coherent state
\cite{QKT}. In a parameter regime corresponding to chaotic dynamics
\cite{Haake}, RMT predicts the long-time asymptotic state of the top
to be described by a random pure state uniformly drawn according to
the Haar measure on SU($N$).  By the above Theorem, the expected
$\mathfrak{su}(2)$-purity may then be estimated as
$$ \overline{P}_{\mathfrak{su}(2)} = \kappa_{\mathfrak{su}(2)}
\frac{3}{N+1} = \frac{1}{2J}.$$ This coincides with the result
obtained in \cite{Kus} by direct integration, and is in excellent
agreement with numerical simulations \cite{QKT}.

\vspace*{1mm}

As noticed, for states obeying an appropriate anti-unitary
symmetry, the components may be chosen real without loss of
generality.  For random states with purely {\em real} components,
only $(N-1)(\frac{N}{2}+1)$ operators are required to span the
space of real traceless observables, resulting in
$${\mathbb E}\{
\bra{\psi}b_i\ket{\psi}^2 \} = \frac{1}{ \kappa_h (N-1)(N/2+1)} =
\frac{2}{N+2}\,,$$
thereby
\begin{equation}
\overline{P}_h = \kappa_h \frac{2 \,\mbox{dim}(h)}{N+2}
\label{ranrPh},
\end{equation}
where $h$ is now understood as a subspace of purely real observables.

\vspace*{1mm}

{\sc Example 3.} The expected value for the $\tt IPR$ in any given
basis for random states with purely real components may be found by
exploiting the connection between ${\tt IPR}$ and $P_{h_{diag}}$ shown
in Eq.~(\ref{Pdiag}). Since $(N-1)$ basis operators span $h_{diag}$
and $\kappa_{h_{diag}} = {1}/({N+1})$, it follows that
$\overline{P}_{h_{diag}} = {2}/({N+2})$. Thus,
\begin{equation}
\overline{\tt IPR}^{(real)} = \frac{3}{N+2}\,.
\label{expipr}
\end{equation}

\vspace*{1mm}

The result given in Theorem 4.1 may also be extended to situations
where the random states of interest belong to a {\em proper subspace}
$\mathcal{S} \subset \mathcal{H}$ with
$\mbox{dim}(\mathcal{S})=N_S$. In general, care should be taken as the
basis operators ${b_i}$ need not remain traceless and normalized after
projection into ${\cal S}$. Let $\Pi$ be the projector onto ${\cal
S}$.  Then $\Pi b_i\Pi = \alpha_i b'_i + \beta_i \mathbb{I}$, where
$\mbox{tr}(b'_i) = 0$, and $\mbox{tr}({b'}_i^2) = N_S$.  Now ${\mathbb
E}\{ \bra{\psi}\alpha_i b'_i + \beta_i\mathbb{I} \ket{\psi}^2 \}=
\alpha_i^2{\mathbb E}\{\bra{\psi}b'_i\ket{\psi}^2\} + \alpha_i \beta_i
{\mathbb E}\{\bra{\psi}b'_i\ket{\psi}\} +\beta^2$.  But ${\mathbb
E}\{\bra{\psi}b'_i\ket{\psi}\}=0$ since ${\mathbb
E}\{|\langle{n}\ket{\psi}|^2\}$ does not depend on $n$, and $b'_i$ is
traceless. Thus, by Eq.~(\ref{bav}) one finds
\begin{equation}
\overline{P}_{h|\cal S} = \kappa_h \Big( \frac{1}{N_S+1} \sum_i
\alpha_i^2 +\sum_i \beta_i^2\Big). \label{ransub}
\end{equation}

\vspace*{1mm}

{\sc Example 4.} Consider the average local purity for pure states of
the $S_z=0$ subspace ${\cal S}_0$ in the state space of $n$ qubits,
which have real components when expressed in $\big\{\ket{k_z}\big\}$
basis, and are uniformly random with respect the Haar measure on
$\mbox{SO}(N_0)$, $\mbox{dim}({\cal S}_0)=N_0$. The only single-qubit
observables having non-vanishing expectation values for states of this
ensemble are $\sigma_z^{(i)}$. Since each $\sigma_z^{(i)}$ is diagonal
in the $\big\{\ket{k_z}\big\}$ basis, $\Pi\sigma_z^{(i)}\Pi$ is also
diagonal. Furthermore, since every (diagonal) matrix element is either
$+1$ or $-1$, it follows that $\mbox{tr}((\Pi\sigma_z^{(i)}\Pi)^2) =
N_0$. But because there are as many $\big\{\ket{k_z}\big\}$ basis
states spanning ${\cal S}_0$ for which the $i$-th qubit is 0 as 1, it
also follows that $\mbox{tr}(\Pi\sigma_z^{(i)}\Pi) = 0$.  Thus, the
local purity of a typical real pure state averaged over the Haar
measure is
$$ \overline{P}_{loc|{\cal S}_0} = \kappa_{loc} \frac{2n}{N_0+2} =
\frac{2}{N_0+2}.$$ \noindent

\vspace*{1mm}

{\sc Example 5.} A similar method may be followed to obtain the
expected purity with respect to other subalgebras of qubit
observables, in particular algebras corresponding to all observables
on selected pairs or $q$-dimensional blocks of spins (e.g. bi-local
purity $P_2$, tri-local purity $P_3$, and so on). Consider, for
instance, the case $q=2$, which is relevant to the analysis in
\cite{Monta}.  That is, we wish to compute $\overline{P}_2$, over pure
states of the $S_z=0$ subspace of an $n$-qubit space, with real
components in the $\big\{\ket{k_z}\big\}$ basis, which are uniformly
random with respect the Haar measure on $\mbox{SO}(N_0)$. Since $P_2 =
\frac{2}{L} \sum_i P_{bl_i}$, where $P_{bl_i}$ is the purity if the
$i$-th 2-qubit block, it suffices to calculate
$\overline{P}_{bl_i}$. The only two-qubit Pauli operators which have
non-zero expectation values for this ensemble are: $\sigma_z^{(1)}$,
$\sigma_z^{(2)}$, $\sigma_z^{(1)} \sigma_z^{(2)}$ , $\sigma_x^{(1)}
\sigma_x^{(2)}$, and $\sigma_y^{(1)} \sigma_y^{(2)}$. The trace and
trace-norm of the projection of each operator into ${\cal S}_0$ may be
found using combinatorial arguments presented in \cite{next}, yielding:
$\mbox{tr}(\Pi\sigma_z^{(1)}\Pi) = \mbox{tr}(\Pi\sigma_z^{(2)}\Pi) =
0$, $\mbox{tr}((\Pi\sigma_z^{(1)}\Pi)^2) =
\mbox{tr}((\Pi\sigma_z^{(2)}\Pi)^2) = N_0$, $\mbox{tr}(\Pi
\sigma_z^{(1)} \sigma_z^{(2)}\Pi)= \sum_{k=0}^{k=2} (-1)^k {2\choose
k} {n-2 \choose n/2-k} = \lambda$, $\mbox{tr}((\Pi \sigma_z^{(1)}
\sigma_z^{(2)}\Pi)^2) = N_0$, $\mbox{tr}(\Pi\sigma_x^{(1)}
\sigma_x^{(2)}\Pi) = \mbox{tr}(\Pi\sigma_y^{(1)} \sigma_y^{(2)}\Pi)=
0$, and $\mbox{tr}((\Pi\sigma_x^{(1)} \sigma_x^{(2)}\Pi)^2) =
\mbox{tr}((\Pi\sigma_y^{(1)} \sigma_y^{(2)}\Pi)^2)= {n-2 \choose
(n-2)/2}$. The coefficients $\alpha_i$ and $\beta_i$ for the traceless
and identity components of the projection of each operator into $S_0$
may be determined from these values.  Thus, applying
Eq.~(\ref{ransub}) finally yields
$$\overline{P}_{2|{\cal S}_0} =
\frac{1}{3}\Big{[}\frac{2}{N_0+2}\Big{[}3 - \frac{\lambda^2}{N_0}
+ \frac{4}{N_0} {L-2 \choose (L-2)/2} \Big{]} +
\frac{\lambda^2}{N_0}\Big{]}.$$

\section{Application to disordered quantum spin chains}

A natural testbed for the above considerations is the study of
many-body quantum systems. Here, we focus on investigating the
relationship between local purity and ${\tt NPC}$ in the eigenstates
of a disordered Heisenberg spin chain across a transition from quantum
integrability to quantum chaos.

Quantum chaos is generally understood as referring to manifestations
of classical chaos at the quantum level. Foremost among these is the
distribution of energy level spacings. As it is by now well
established, classically integrable (chaotic) systems typically
exhibit a Poisson (Wigner-Dyson) level statistics distribution
\cite{Haake}. For systems without an obvious classical counterpart,
for instance spin chains, the presence of a Poisson or Wigner-Dyson
level spacing distribution is taken as a phenomenological criterion
for labelling the system as integrable or, respectively, chaotic.

In what follows, we shall consider a representative disordered quantum
spin $1/2$ system within a class of Heisenberg models in a transverse
field which we discuss in full generality in \cite{next}. In
particular, we choose a one-dimensional quantum spin chain described
by the following Hamiltonian:
\begin{eqnarray}
H = H_0 + H_{int} =
\sum_{i=1}^{n}\frac{\varepsilon_i}{2}\sigma_z^{(i)}
+\frac{J}{4} \sum_{i=1}^{n-1}
\vec{\sigma}^{(i)}\cdot\vec{\sigma}^{(i+1)}, \label{Heis}
\end{eqnarray}
where $\varepsilon_i=\varepsilon + \delta\varepsilon_i$, $\varepsilon$
and $J$ are fixed positive numbers, $\delta\varepsilon_i$ are uniform
random variables within the interval $[-d/2,d/2]$, and open boundary
conditions are imposed. Because $H$ commutes with the $z$-component of
the total spin angular momentum $S_z$, each invariant subspace may be
diagonalized independently.  We focus on the band with no net
magnetization, the $S_z = 0$ subspace.  Eigenvalues and eigenvectors
have been computed numerically for chains of size up to $n=12$. This
yields $N = {12 \choose 6} = 924$ as the dimension of the relevant
$S_z=0$ subspace.

When $J/d = 0$, $H=H_0$ is trivially solvable, and for
sufficiently small $J/d$ perturbation theory is valid.  In this
regime, the system has Poisson level statistics. When $d\sim J$,
perturbation theory breaks down, and a cross-over from Poisson to
Wigner-Dyson level statistics occurs.

\begin{figure}[b]
\begin{center}
\includegraphics[width=3.4in]{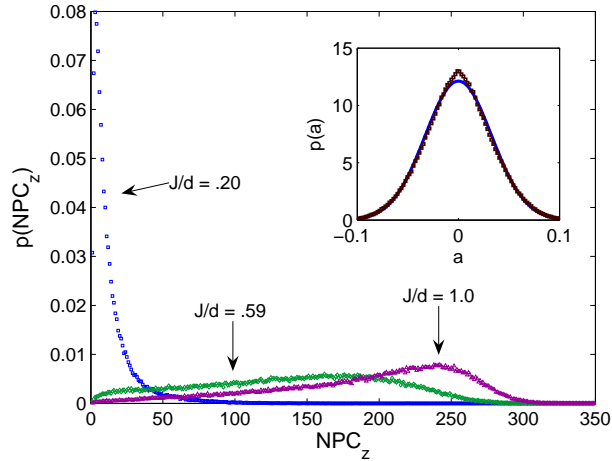}
\caption{Eigenvector distribution versus ${\tt NPC}_z$ at $J/d$ =
  0.20, 0.59, and 1.0, for model Hamiltonian (\ref{Heis}) with $n=12$
  spins. Inset: Distribution of eigenvector components, $a$, for
  eigenvectors with $300<{\tt NPC}_z<316$, based on 300 random
  realizations at $J/d = 1$. The smooth curve is a Gaussian
  distribution with $\sigma^2 = {1}/{924}$. } \label{iprh}
\end{center}
\end{figure}

Associated with the transition in level statistics, there is a
transition from eigenvectors which are well localized in the
eigenbasis of $H_0$, to eigenvectors which are delocalized and
approximately random (Fig.~\ref{iprh}). Generally, for fully developed
chaos, the eigenvectors achieve a distribution of components uniform
over the surface of an $N$-sphere.  In systems obeying time-reversal
invariance or, more generally as mentioned, an appropriate
anti-unitary symmetry \cite{Haake,Berko}, this is equivalent to a
Gaussian distribution of eigenstate components in the limit of large
$N$. However, as seen in Fig.~\ref{iprh}, for this model the states
for which ${\tt NPC_z}$ is near the expected value for random states
of $({N+2})/{3}$ \cite{caveat2} have a component distribution which is
only approximately Gaussian.  Furthermore, there is no regime where
most eigenvectors have an ${\tt NPC_z}$ consistent with the expected
value for random states, although this value does serve as an
approximate upper bound on delocalization. At specific $J/d$ values,
this model typically exhibits a fairly wide distribution of ${\tt
NPC_z}$.

\begin{figure}[h]
\begin{center}
\includegraphics[width=3.4in]{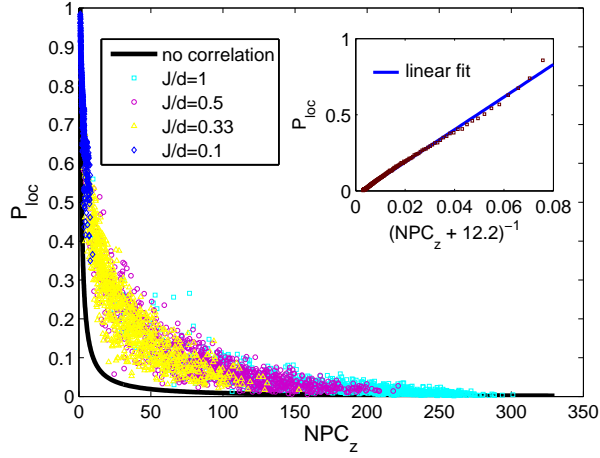}
\caption{$P_{loc}$ versus ${\tt NPC}_z$ for each eigenvector in the
central band using a single random realization at four different
values of J/d.  Inset: Linear fit of $P_{loc}$ to $({\tt
NPC}_z+12.2)^{-1}$ over all eigenvectors and 100 random realizations
at $J/d = 0.59$.}
\label{pvi}
\end{center}
\end{figure}

Throughout the localized-to-delocalized transition, we examined
the relationship between ${\tt NPC}_z$ and local purity for each
eigenvector in the $S_z=0$ subspace. In Fig.~\ref{pvi},
$P_{loc}$ is plotted against ${\tt NPC}_z$ for each eigenvector
using a single random disorder realization and four representative
values of $J/d$. At $J/d = 0.59$, $P_{loc}$ is averaged over each
eigenvector between integer values of ${\tt NPC}_z$ and over 100
disorder realizations, resulting in a smooth curve which closely
fits
\begin{equation}
P_{loc} (|\psi\rangle) = \frac{14.5}{{\tt NPC}_z (|\psi\rangle) +
12.2} - 0.032, \label{NumDat}
\end{equation}
see inset in Fig.~\ref{pvi}. The value $J/d = 0.59$ is chosen because
of the corresponding wide distribution of ${\tt NPC}_z$.  Similarly
constructed average curves for other values of $J/d$, however, do not
show significant differences.

\begin{figure}[h]
\begin{center}
\includegraphics[width=3.4in]{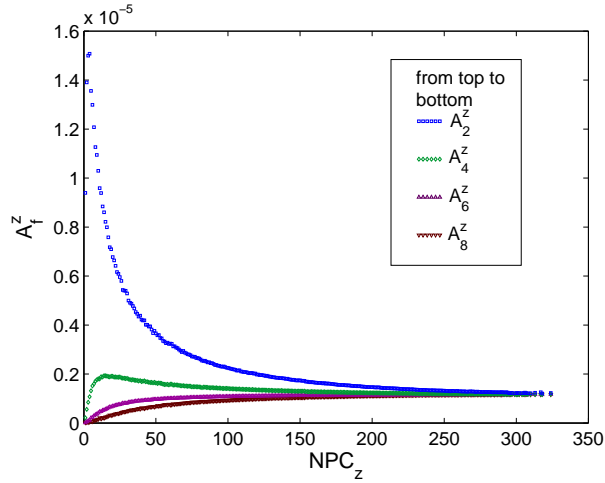}
\caption{Dependence of averages $A^z_f$ on ${\tt NPC}_z$ at $J/d =
  0.59$. Each $A^z_f$ is averaged between integer values of ${\tt
    NPC}_z$ and over all eigenvectors and 100 random realizations.}
\label{avi}
\end{center}
\end{figure}

In spite of qualitative agreement, the averaged relationship between
$P_{loc}$ and ${\tt NPC}_z$ given in Eq.~(\ref{NumDat}) deviates from
the relationship predicted in Eq.~(\ref{no dep Sz0}) under the
assumption of Hamming uncorrelation between $A_f^z =\overline{|a_k^z|^2
|a_j^z|^2}$ and $f_{kj}$. This indicates that the $A_f^z$ {\em do}
depend non-trivially on Hamming distance in general. In
Fig.~\ref{avi}, $A_f^z$ is averaged over each eigenvector between
integer values of ${\tt NPC}_z$ and 100 disorder realizations at $J/d
= 0.59$.  Especially for small ${\tt NPC}_z$, $A_f^z$ tends to be
larger for smaller values of $f_{kj}$. There is a strong peak in
$A_2^z$ at ${\tt NPC}_z \approx 4$, and a less pronounced peak in
$A_4^z$ at ${\tt NPC}_z \approx 15$. All other $A_f^z$ rise gradually.
As ${\tt NPC}_z$ approaches the limiting value of $(N+2)/3$, the
values of the $A_f$ become closer to each other, and appear in good
agreement with the expected value for random states,
${(N-3)}/{[N(N-1)(N+2)]}$ \cite{Af}. $A_{10}^z$ and $A_{12}^z$ (data
not shown) lie close to $A_{8}^z$.

In the perturbative regime, the dependence of $A_f^z$ on Hamming
distance $f_{kj}$ may be understood as a consequence of the two-body
form of the interaction.  For small $J/d$, the eigenvectors may be
expanded in a perturbation series. Starting with an arbitrary
eigenvector $\ket{k}$ of $H_0$,
$$\ket{k} \mapsto \ket{k} + \sum \frac{J}{E_k - E_j}\ket{j} + \sum
\frac{J^2}{(E_k-E_j)(E_k-E_l)}\ket{l} + \ldots $$
\noindent
The diagonal contribution $\sigma_z^{(i)}\sigma_z^{(i+1)}$ may be
incorporated into the unperturbed diagonal energies for the current
reasoning. Since the Heisenberg interaction only couples eigenstates
of $H_0$ which are Hamming distance of $2$ away from each other, every
state $\{\ket{j}\}$ that appears in the first-order sum is a Hamming
distance of 2 away from $\ket{k}$.  The states $\{\ket{l}\}$ that
appear in the second-order sum are Hamming distance 2 or 4 away from
$\ket{k}$.  Thus, for an eigenstate in the perturbative regime, all of
the first-order products will contribute to $A_2^z$.  No product
larger than second-order will contribute to $A_4^z$, and so on.  After
the breakdown of perturbation theory, the $A_f^z$ continue to show a
dependence on order in perturbation theory for all values of ${\tt
NPC_z}$, however, the effect decreases as $\tt NPC_z$ approaches the
random state value of $({N+2})/{3}$.

An interesting question is the behavior of the relationship between
${\tt NPC_z}$ and $P_{loc}$ in the thermodynamic limit where
$n\rightarrow \infty$. Because the relationship between local purity
and ${\tt NPC_z}$ involves averages over all pairs of basis states of
fixed Hamming distance, it is reasonable to conjecture that the
relationship should become increasingly sharp in this limit, provided
that the average value of $|a_k|^2|a_j|^2$ for fixed order in
perturbation theory exists.

\section{Conclusion}

We have quantified the relationship between delocalization as measured
by ${\tt NPC}$ in a maximal set of mutually unbiased product bases and
global entanglement as measured by local purity.  In general, the
relationship between the two depends on how products of state vector
components are {\em correlated with respect to Hamming distance} -- or
a suitable generalization for higher-dimensional subsystems. Under the
condition that no such correlation exists, a simple relationship
between ${\tt NPC}$ in each basis and local purity is established. For
states with certain physically relevant symmetries, the number of
bases may be reduced. In addition, for each state, there always exists
a basis in which the local purity is related to ${\tt NPC}$ in this
single basis through correlations with respect to Hamming distance.
Such analysis yields an expression for local entanglement, GE$_{loc}$,
as the expectation value of Hamming distance between measurements of
two copies of the same pure state in the state-dependent canonical
basis where each reduced density matrix is diagonal.

Distributions of random states under which the assumption of
uncorrelation is naturally satisfied are also discussed.  A simple
method to calculate the expected relative purity over an ensemble of
pure states invariant under the Haar measure is introduced, and
illustrated in several examples.  Lastly, the connection between local
purity and correlations between products of components is investigated
numerically for a disordered Heisenberg spin chain.  Because the
deviation of the relationship between $P_{loc} $ and ${\tt NPC}_z$
from that predicted under the uncorrelation assumption is likely a
consequence of the two-body nature of the interaction, a similar
relationship is predicted to hold for any disordered qubit system with
two-body interactions which has symmetry properties allowing ${\tt
NPC}$ in a single basis to enter the relationship with $P_{loc}$.  For
systems without such symmetries, we conjecture that the $P_\alpha$
associated with ${\tt NPC}$ in the eigenbasis of $H_0$ will still
provide the main contribution to $P_{loc}$, until the eigenvectors are
maximally random.  Thus, the relationship between $P_{loc}$ and ${\tt
NPC}$ in the eigenbasis of $H_0$ may be generic to all disordered
many-body systems.

As a general remark, we also expect the correlation between products
of components and Hamming distance to be relevant to other
entanglement measures.  For instance, the $n$-tangle is written as a
sum of products of pairs that are Hamming distance $n$-apart.  Thus,
we conjecture that this characteristic structure may be important for
the study of entanglement properties across a localized-to-delocalized
transition and across quantum criticality in many-body systems.

\section*{Acknowledgments}

L.V. is especially indebted to Howard Barnum and Gerardo Ortiz for
continuous exchange and uncountable discussions on the meaning and
usefulness of generalized entanglement.  The authors also thank Simone
Montangero, Lea F. Santos, and Yaakov S. Weinstein for feedback and a
critical reading of the manuscript, and Jay Lawrence for discussions
on mutually unbiased bases. W.G.B. gratefully acknowledges partial
support from Constance and Walter Burke through their Special Projects
Fund in Quantum Information Science, and current support from a GAANN
Fellowship.


\begin{thebibliography}{10}
\bibitem{remark} Note that such a projection need not lead to a
positive (semidefinite) operator ${\cal P}_h(\rho)$ on the full state
space.  If the identity operators is included in $h$, ${\cal
P}_h(\rho)$ is positive in the standard multipartite case and positive
in general upon restricting its domain to $h$ \cite{Cones,VB}.  In
practice, since omitting the identity has no effect on the convex
structure, traceless distinguished observables are often considered
to the purposes of constructing measures of GE.
\bibitem{caveat} Note that $P_z$ should not be confused with
$P_{diag}$, which involves in this case expectations of arbitrary
(non-local) strings of $\sigma_z$ operators on different qubits.
\bibitem{caveat2} Strictly speaking, $(N+2)/3$ is the expected value
of {\tt IPR}, cf Eq.~(\ref{expipr}).  However, because the
distribution of states with values around the expected {\tt IPR}
becomes quickly sharp as $N$ increases, we identify $\overline{\tt
NPC}\approx {\overline{\tt IPR}}^{-1}$.
\bibitem{Af} This value is obtained by using $\overline{|\langle
\psi|\psi\rangle|^2} = \sum_n \overline{a_n^4} + \sum_{n,m}
\overline{a_n^2 a_m^2} = 1$ (where $z$-dependences are implicit
throughout).  As $\overline {a_n^2 a_m^2}$ does not depend on $n$ and
$m$, and $\sum _n \overline{a_n^4} = \overline{\tt IPR} =
{3}/({N+2})$, it follows that ${3}/({N+2}) + N(N-1)\,\overline{a_0^2
a_1^2} = 1$.  Identifying $\overline{A}_f$ with $\overline{a_0^2
a_1^2}$ and rearranging terms yields the result quoted in the text.
\end{thebibliography}
\end{document}